\documentclass[9pt,twocolumn,twoside]{osajnl}

\journal{ol} 

\setboolean{shortarticle}{true}

\title{Spectrally pure photon pair generation in asymmetric heterogeneously coupled waveguides}

\author[1]{Xiangyan Ding}
\author[1]{Jing Ma}
\author[1]{Liying Tan}
\author[2]{Amr S. Helmy}
\author[1,*]{Dongpeng Kang}

\affil[1]{School of Astronautics and National Key Laboratory of Science and Technology on Tunable Laser, Harbin Institute of Technology, 92 West Dazhi Street, Harbin, 150001, China}
\affil[2]{The Edward S. Rogers Department of Electrical and Computer Engineering, University of Toronto, 10 King’s College Road, Toronto, Ontario, Canada M5S 3G4}

\affil[*]{Corresponding author: dongpeng.kang@hit.edu.cn}

\begin{abstract}
In this work, we develop a design methodology to generate spectrally pure photon pairs in asymmetric heterogeneously coupled waveguides by spontaneous parametric down-conversion. Mode coupling in a system of waveguides is used to directly tailor the group velocity of a supermode to achieve group velocity matching (GVM) that is otherwise not allowed by material dispersion. Design examples based on thin film lithium niobate waveguides are provided, demonstrating high spectral purity and temperature tunability. This approach is a versatile strategy applicable to waveguides of different materials and structures, allowing more versatility in single-photon source designs. 
\end{abstract}

\setboolean{displaycopyright}{true}

\begin{document}

\maketitle

Single photons are a vital resource in various applications of quantum information processing, including quantum key distribution, optical quantum computing, simulation and metrology, etc\cite{flamini2019photonic}. The underline quantum processes that rely on the interference between different quantum states of light require high purity and indistinguishability of these states. An ideal approach to generate single photons is to use quantum emitters such as quantum dots \cite{utzat2019coherent} or crystal vacancies \cite{ISI:000526348400042}, which emit one photon at a time. Although tremendous progress has been made in this approach, it is still a challenge to generate indistinguishable photons from different emitters at different wavelength regimes\cite{Baoeaba8761}. Moreover, such quantum emitters often do not afford room-temperature operation\cite{Seravalli_2020} and require expensive cryogenic cooling. A more readily available approach that has been widely used in various quantum optical experiments \cite{slussarenko2019photonic} is to use heralded single-photon sources to generate photon pairs via nonlinear optical processes including spontaneous parametric down-conversion (SPDC) in $\chi^{\left(2\right)}$ and spontaneous four-wave mixing (SFWM) in $\chi^{\left(3\right)}$ nonlinear media \cite{Hashimoto:21,chen2019indistinguishable,christensen2018engineering}. In such a case, the detection of one photon indicates the existence of its twin \cite{doi:10.1063/5.0030258}.

For heralded single photons to be in a pure quantum state, paired photons should not be correlated in frequency or any other degree of freedom. However, this is usually not the case, as the purity of heralded single photons generated by SPDC are often limited by the spectral correlations between photons in a pair due to energy conservation. A common strategy is to use narrow-band spectral filters to remove these correlations, which, however, is obtained at the expense of heralding efficiency and increased source complexity. While micro-resonators which provide enhanced nonlinear interactions can be used to remove spectral correlations\cite{Burridge:20}, for typical non-resonant waveguides, group velocity matching (GVM) \cite{grice2001eliminating,Wei:21} needs to be employed in order to directly generate uncorrelated photon pairs on-chip. For example, material dispersion in potassium titanyl phosphate (KTP) naturally satisfies GVM in a certain wavelength range, thus allowing the generation of pure heralded single photons in periodically poled KTP (PPKTP) waveguides in a given wavelength range\cite{Padberg:20,Kaneda:20}. However, material dispersion in most materials do not satisfy GVM requirements, and thus do not allow the direct generation of pure heralded single photons.

On the other hand, dispersion engineering in integrated photonic chips provides a powerful tool to achieve functions that are otherwise not achievable, and has been applied in a broad range of applications such as parametric optical processes \cite{Nitiss:20} and supercontinuum generation \cite{gonzalez2019design}, etc. Several photonic structures including Bragg gratings \cite{doi:10.1063/5.0022963}, nonlinear photonic nanowires \cite{XU201735}, and asymmetric slot waveguides \cite{kim2016broadband} have been adopted to engineer the dispersion. 

In this work, we develop a technique to achieve GVM by dispersion engineering in asymmetric heterogeneously coupled waveguides. Instead of focusing on group velocity dispersion (GVD) as in previous work \cite{peschel1995compact,mia2019extremely}, we utilize a mode-coupling approach to tailor the group velocities of the modes involved to satisfy GVM in materials that are otherwise unachievable. This is a generic technique that can be used in nonlinear waveguides of different materials to achieve GVM that is not naturally allowed by material dispersion. As an illustrative example, we provide structures based on asymmetric heterogeneously coupled lithium niobate (LN)- arsenic selenide (As$_2$Se$_3$) that are capable of generating spectral pure photon pairs. 

The two-photon state generated by SPDC in a nonlinear waveguide is given by\cite{PhysRevA.56.1627}
\begin{equation}
\vert\psi\rangle=\int d\omega_S\int d\omega_If(\omega_S,\omega_I)\vert\omega_S\rangle\vert\omega_I\rangle,
\label{Eq:two-photon state}
\end{equation}
where $\omega_S$ and $\omega_I$ are the frequencies of signal and idler photons, respectively, and $f(\omega_S,\omega_I)$ is the normalized bi-photon wave function (BFW) given by $\alpha(\omega_S + \omega_I)\phi(\omega_S,\omega_I)$. Here $\alpha(\omega_S + \omega_I)$ is the pump spectral amplitude and $\phi(\omega_S,\omega_I)$ is the phase-matching function, usually in the form of a sinc function. For pure heralded single photons, the BWF must be factorable, i.e., $f(\omega_S,\omega_I) = f_S(\omega_S)f_I(\omega_I)$, with $f_S(\omega_S)$ and $f_I(\omega_I)$ being two single-variable functions. To achieve this directly without spectral filtering, the group velocities of the pump and down-converted photons need to satisfy the following condition of GVM \cite{grice2001eliminating}
\begin{equation}
v_S\leq v_P\leq v_I, \mathrm{or}\; v_I\leq v_P\leq v_S.
\label{Eq:GVM}
\end{equation}
However, this condition is not satisfied except for a handful of cases. Typically, $v_P$ is significantly lower than both $v_S$ and $v_I$.

\begin{figure}[bt!]
	\centering\includegraphics[width=\columnwidth]{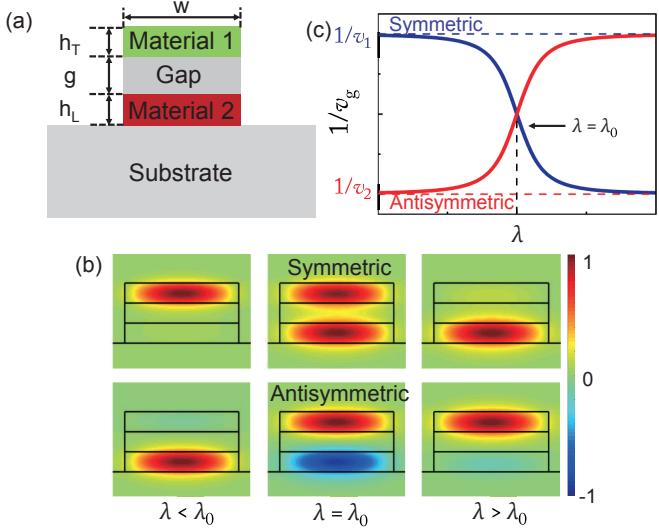}
	\caption{(a) Schematic cross-sectional view of an asymmetric heterogeneously-coupled waveguide with a mode crossing point at $\lambda_0$. (b) Modal field profiles of symmetric (upper row) and antisymmetric (lower row) supermodes at different wavelengths of $\lambda<\lambda_0$, $\lambda=\lambda_0$, and $\lambda>\lambda_0$. (c) The inverse group velocities of the symmetric (blue) and antisymmetric (red) supermodes given by Eq. (\ref{Eq:supermode vg}).}
	\label{fig1}
\end{figure}

Here we consider an asymmetric heterogeneously coupled waveguide with a structure schematically shown in Fig. \ref{fig1} (a), in which two waveguides of different materials are placed close enough such that the waveguide mode of one couples to the other. Fig. \ref{fig1} (b) gives the modal electric field profiles of symmetric (upper row) and antisymmetric (lower row) supermodes at different wavelengths $\lambda$, where $\lambda_0$ is the mode crossing wavelength. For a coupling coefficient $\kappa$ that depends on the waveguide spacing and index difference, the propagation constants of the coupled symmetric ($\beta_+$) and antisymmetric ($\beta_-$) supermodes are given by \cite{peschel1995compact}
\begin{equation}
\beta_\pm=\frac{1}{2}\left(\beta_1+\beta_2\right)\pm\sqrt{\left(\frac{\beta_1-\beta_2}{2}\right)^2+\vert\kappa\vert^2},
\label{Eq:supermodes}
\end{equation}
where $\beta_{i}$ ($i=1,2$) is the propagation constant of an isolated mode. It's a function of  frequency $\omega$, and by neglecting higher-order dispersions, can be expressed as $\beta_{i}\left(\omega\right)\approx\beta_{i}\left(\omega_0\right)+(\omega-\omega_0)/v_{i}$, with $v_i=d\omega/d\beta_i$ being the modal group velocity at the mode crossing point $\omega_0$. By substituting the approximate values of $\beta_1$ and $\beta_2$ given above into Eq. (\ref{Eq:supermodes}) and performing first-order derivative with respect to $\omega$, the group velocities of the coupled supermodes can be obtained as 
\begin{equation}
\frac{1}{v_\pm(\omega)}=\frac{1}{2}\left(\frac{1}{v_1}+\frac{1}{v_2}\right)\pm\frac{1}{2}\left(\frac{1}{v_1}-\frac{1}{v_2}\right)\widetilde{\omega}\left(\widetilde{\omega}^2+1\right)^{-1/2},
\label{Eq:supermode vg}
\end{equation}
where $\widetilde{\omega}=\left(\omega-\omega_0\right)/\delta\omega$, with $\delta\omega$ being the characteristic bandwidth defined by
\begin{equation}
\delta\omega=2\vert\kappa\vert\left|\frac{1}{v_1}-\frac{1}{v_2}\right|^{-1}.
\end{equation}
In obtaining Eq. (\ref{Eq:supermode vg}), we have neglected the frequency dependence of $\kappa$, which is small in the frequency range of interest.

This group velocity tunability offered by waveguide coupling is illustrated in Fig. \ref{fig1}(c). In the extreme case of $\widetilde{\omega}\rightarrow\infty$, the inverse of group velocities can be obtained as $1/v_+\rightarrow1/v_1$ and $1/v_-\rightarrow1/v_2$. In the opposite extreme of $\widetilde{\omega}\rightarrow-\infty$, they are given by $1/v_+\rightarrow1/v_2$ and $1/v_-\rightarrow1/v_1$ instead. These results are intuitively understandable, because in either case, the modal field locates mostly in one of the waveguides, as shown in Fig. \ref{fig1}(b), and therefore the modal dispersion of a coupled supermode is dominated by a that of single waveguide. As a result, the group velocities of two isolated waveguide modes provide the upper and lower bounds for those of the coupled supermodes.

This strategy can be applied to tailor the group velocity of one of the down-converted photons, such that it is generated in a supermode with a center wavelength in the coupling regime. By choosing an appropriate combination of material and dimensions for the neighboring waveguide and let it couple with the original one, GVM could be satisfied at the desired wavelength. It should be emphasized that this is a versatile technique applicable to waveguides of different materials and structures. Although Fig. \ref{fig1}(a) shows a vertically coupled ridge waveguide, in reality, coupled waveguides of any geometry could be used, depending on the available fabrication techniques associated with materials of consideration.

\begin{figure}[tb!]
	\centering\includegraphics[width=\columnwidth]{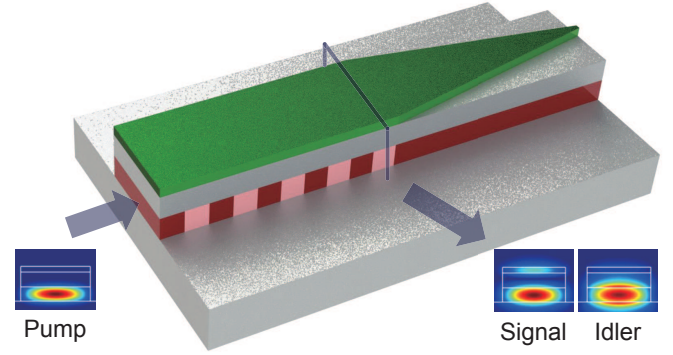}
	\caption{Schematic of a vertically coupled As$_2$Se$_3$-LN-on-insulator waveguide and the related modal intensity profiles, with LN, SiO$_2$ and As$_2$Se$_3$ being colored in red, gray and green, respectively. An adiabatic taper could be used on the As$_2$Se$_3$ layer after the QPM section to convert the supermode of the coupled waveguide to the fundamental mode of a LN-on-insulator waveguide.}
	\label{fig2}
\end{figure}

To demonstrate the proposed technique, here we provide a design example based on $z$-cut $y$-propagating thin film LN-on-insulator waveguides, in which the TE polarized pump at 775 nm generates cross-polarized photon pairs centered around 1550 nm via the type-II SPDC process using quasi-phase matching (QPM). The LN film is bonded on SiO$_2$ and is assumed to have a width $w_L=2$ \textmu m and thickness $h_L=0.48$ \textmu m. The group velocities of pump, signal (assumed to be TE) and idler (assumed to be TM) are calculated to be $v_P=1.23\times10^8$ m/s, $v_S=1.27\times10^8$ m/s and $v_I=1.26\times10^8$ m/s. Thus our goal is to reduce $v_S$ so that $v_S\leq v_P$ while keeping $v_I$ unattended. 

Here we choose As$_2$Se$_3$ \cite{zhang2014nonlinear} as the coupling waveguide material, because of its faster index variation than LN that allows a lower group velocity in an isolated waveguide, and a layer of lower index SiO$_2$ is used as the spacing. The structure and modal profiles are schematically shown in Fig. \ref{fig2}. The signal photon is designed to be generated in the TE polarized antisymmetric supermode, with its group velocity tailored by the thicknesses of As$_2$Se$_3$ layer ($h_T$) and SiO$_2$ layer ($g$). After photon pairs are generated, an adiabatic taper on the As$_2$Se$_3$ layer could be added to convert the supermode to the fundamental mode of the isolated LN waveguide to minimize the loss occurs during coupling with the following photonic circuit or fiber.

\begin{figure}[tb!]
	\centering\includegraphics[width=\columnwidth]{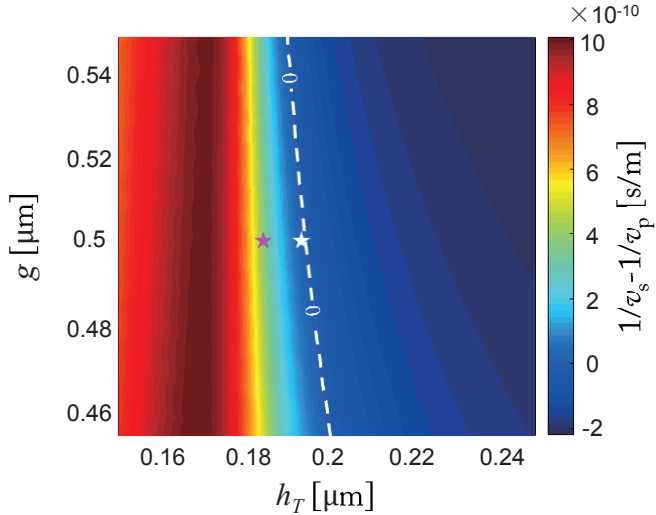}
	\caption{Difference between the inverse group velocities of the signal photon and the pump photon, $1/v_S-1/v_P$, as a function of the thicknesses of As$_2$Se$_3$ layer ($h_T$) and SiO$_2$ layer ($g$). The contour represents $v_S=v_P$.}
	\label{fig3}
\end{figure}

Fig. \ref{fig3} shows the difference between the inverse group velocities of the signal and pump photons, $1/v_S-1/v_P$, as a function of waveguide geometric parameters $h_T$ and $g$. The contour line with a value of $0$ represents $v_S=v_P$. Any combination of $h_T$ and $g$ that falls on or to the left of this line ensures GVM is satisfied. Two specific designs, labeled by white and purple marks in Fig. \ref{fig3}, are discussed in the following.

\begin{figure}[ht!]
	\centering\includegraphics[width=\columnwidth]{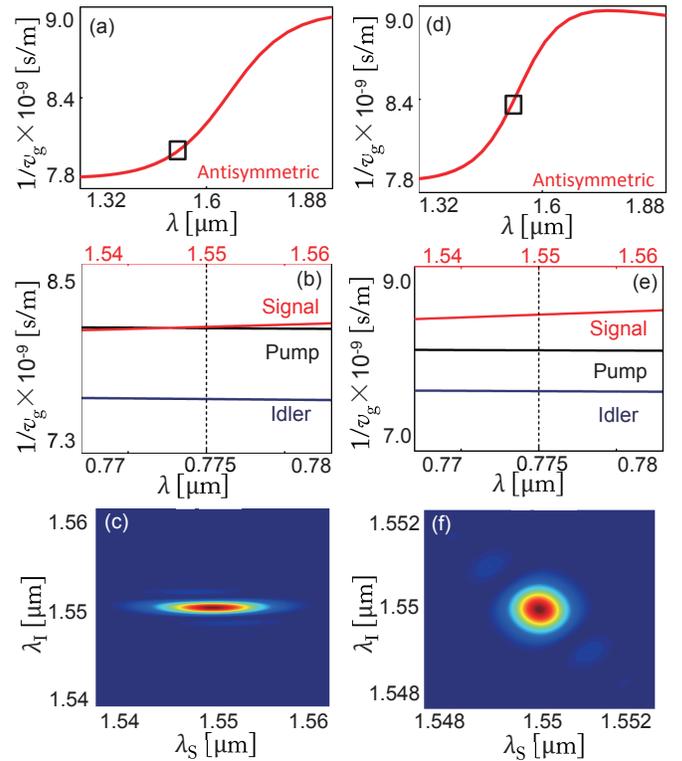}
	\caption{(a) The inverse group velocity of the antisymmetric supermode of the design example with $h_T=193$ nm and $g=500$ nm, and (b) inverse group velocities of the pump, signal, and idler near the phase matching wavelength, and (c) JSI of the photon pairs. (d) The inverse group velocity of the antisymmetric supermode of the design example with $h_T=184$ nm, $g=500$ nm, and (e) inverse group velocities of the pump, signal, and idler near the phase matching wavelength, and (f) JSI of the photon pairs. Markers in (a) and (d) represent the working wavelength.}
	\label{fig4}
\end{figure}

In the first case, with $h_T=193$ nm and $g=500$ nm, the inverse group velocity of the antisymmetric supermode as a function of wavelength is shown in Fig. \ref{fig4}(a). Group velocities of the signal and pump photons are almost identical at the signal wavelength of 1550 nm, i.e., $v_S=v_P$, as shown in Fig. \ref{fig4}(b). The joint spectral intensity (JSI), defined by $\vert f(\omega_S,\omega_I)\vert^2$, is shown in Fig. \ref{fig4}(c), where we take Gaussian pump pulses with a duration of 500 fs, and the waveguide length of $L=16$ mm.

To quantify the degree of spectral entanglement, Schmidt decomposition $f\left(\omega_S,\omega_I\right)=\sum_n\sqrt{p_n}u_n(\omega_S)v_n(\omega_I)$ is performed, where $p_n$ is the probability of obtaining the $n$--th state, with $\sum_np_n=1$. The degree of factorizability can be quantified by the Schmidt number $K=1/\sum_np_n^2$. For the JSI shown in Fig. \ref{fig4} (c), the corresponding Schmidt number is $K=1.05$, which is  close to 1, proving that the two-photon state is almost factorizable. The generation rate is calculated to be $\eta=3.4\times10^{-8}$ pairs/pump photon according to the theory in \cite{zhukovsky2012bragg}, assuming a first order QPM grating is used. This value is a factor of 4 smaller than that of a comparative LN-on-insulator waveguide without the As$_2$Se$_3$ layer, indicating the nonlinear overlap does not decrease significantly. Note that the remaining spectral correlation due to the side lobes of a JSI can be further removed with the adoption of Gaussian phase matching \cite{BenDixon:13}.

In the second case with $h_T=184$ nm and $g=500$ nm, the inverse group velocity of the antisymmetric supermode as a function of wavelength is shown in Fig. \ref{fig4}(d). The signal group velocity is further decreased such that $2/v_P\approx1/v_S+1/v_I$, as shown in Fig. \ref{fig4}(e). With a pump pulse duration of 3.5 ps, the JSI is circularly shaped, as shown in Fig. \ref{fig4}(f). The Schmidt number in this case is $K =1.17$, and the generation rate is $\eta=9.2\times10^{-9}$ pairs/pump photon.

In both cases, the modal overlap is assumed to be constant within the frequency range of interest in calculating the BWF and generation rate. This is because mode couplings take place within a wavelength range of a few hundred nanometers, according to Figs. \ref{fig4}(a) and (c), significantly larger than the spectral bandwidths of the photon pairs. It is also in sharp contrast to similar structures in \cite{peschel1995compact,mia2019extremely}, where extremely high GVDs are obtained within bandwidths of only a few nanometers.

\begin{figure}[tb!]
	\centering\includegraphics[width=\columnwidth]{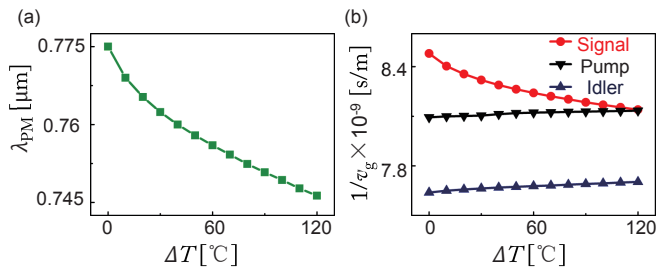}
	\caption{(a) The dependence of phase matching wavelength $\lambda_{PM}$ on temperature variation for the design with $h_T=184$ nm and $g=500$ nm. (b) The  corresponding temperature dependences of inverse group velocities of the three modes.}
	\label{fig5}
\end{figure}

An important feature of these waveguides is that they could be thermally tuned such that the shape of the JSI is switched from circular to elliptical, or vice versa. Taking the second design discussed above as an example, the phase matching wavelength decreases with the increase of temperature, as shown in Fig. \ref{fig5}(a). This leads to the condition of $v_S=v_P$, shown in Fig. \ref{fig5}(b), and thus an elliptically shaped JSI similar to that of Fig. \ref{fig4}(c). The temperature tuning range required ($\Delta T\approx120~ ^\circ$C in this example) depends on the thermo-optic coefficients ($\mathrm{d}n/\mathrm{d}t$) of all materials involved as well as the initial separation between $v_S$ and $v_I$. For a separate design based on LN-Si hybrid platform, we find the temperature tuning range is reduced to $\sim20~^\circ$C, due to the much larger thermo-optic coefficient of Si ($\sim10^{-4}$ /K) than that of As$_2$Se$_3$ ($\sim10^{-5}$ /K) and a lower $v_I$. As such, given the versatility and hybrid nature of these structures, which are well feasible with existing technologies, the technique and designs proposed provide a viable route for on-chip generation of tunable photon pairs.

We emphasize again that the technique presented here is not limited to any particular material system or waveguide geometry. Furthermore, even QPM is not necessary. One can design a waveguide which utilizes modal phase matching (MPM) similar to that in \cite{luo2019semi-nonlinear}, while simultaneously achieving GVM by dispersion engineering via waveguide coupling. In addition, tailoring of group velocity enabled by this technique can be utilized to other applications where group velocity control is critical, such as broadband wavelength conversion \cite{kim2016broadband,Nitiss:20} and polarization entangled photon generation\cite{kang2014two-photon}.

In summary, we have developed a generic technique to generate spectral pure photon pairs in asymmetric heterogeneously coupled waveguides. By engineering the waveguide structure, the group velocity of a coupled supermode can be tailored such that GVM is achieved. This allows the generation of spectral pure photon pairs without the need for bandpass filtering. Design examples based on vertically coupled As$_2$Se$_3$-LN-on-insulator waveguides are provided. This technique is also directly applicable to the continuously-variable regime\cite{PhysRevLett.116.143601} and applications involving temporal modes\cite{Raymer2020}.

\section*{Funding}
National Natural Science Foundation of China (61705053); China Postdoctoral Science Foundation (2016M600249), Heilongjiang Postdoctoral Special Funds; Fundamental Research Funds for the Central Universities, Natural Sciences and Engineering Research Council of Canada.

\section*{Disclosures}

The authors declare no conflicts of interest.

\bibliography{paper}

\bibliographyfullrefs{paper}

\end{document}